\documentclass[%
reprint,
amsmath,
amssymb,
aps,
prb,
]{revtex4-2}

\usepackage{graphicx}
\usepackage{dcolumn}
\usepackage{bm}
\usepackage{physics}
\usepackage{amsmath}
\usepackage{xcolor}

\begin{document}

\preprint{APS/123-QED}

\title{Absence of topological protection of the interface states in $\mathbb{Z}_2$ photonic crystals}

\author{Shupeng Xu}
\author{Yuhui Wang}%
\author{Ritesh Agarwal}
\email{riteshag@seas.upenn.edu}
\affiliation{
Department of Materials Science and Engineering, University of Pennsylvania,\\Philadelphia, 19104, PA, US}
\date{\today}

\begin{abstract}
Inspired from electronic systems, topological photonics aims to engineer new optical devices with robust properties. In many cases, the ideas from topological phases protected by internal symmetries in fermionic systems are extended to those protected by crystalline symmetries. One such popular photonic crystal model was proposed by Wu and Hu in 2015 for realizing a bosonic $\mathbb{Z}_2$ topological crystalline insulator with robust topological edge states, which led to intense theoretical and experimental studies. 
However, rigorous relationship between the bulk topology and edge properties for this model, which is central to evaluating its advantage over traditional photonic designs, has never been established.
In this work we revisit the expanded and shrunken honeycomb lattice structures proposed by Wu and Hu by using topological quantum chemistry tools and show that they are topologically trivial in the sense that symmetric, localized Wannier functions can be constructed. We show that the $\mathbb{Z}$ and $\mathbb{Z}_2$ type classification of the Wu-Hu model are equivalent to the $C_2T$ protected Euler class and the second Stiefel-Whitney class respectively, with the latter characterizing the full valence bands of Wu-Hu model indicating only a higher order topological insulator (HOTI) phase. We show that the Wu-Hu interface states can be gapped by a uniform topology preserving $C_6$ and $T$ symmetric perturbation, which demonstrates the trivial nature of the interface. Our results reveals that topology is not a necessary condition for the reported helical edge states in many photonics systems and opens new possibilities for interface engineering that may not be constrained to require topological designs.
\end{abstract}

\maketitle

Topological photonics began with the seminal work by Raghu and Haldane \cite{raghu2008analogs,haldane2008possible} where the idea of topology in the electronic band structures were generalized to waves in periodic media, leading the way for realizing topological phenomena in artificial structures \cite{lu2014topological,ozawa2019topological,zhang2018topological}.
After the early explorations of photonic Chern insulators where the time-reversal symmetry is explicitly broken \cite{wang2008reflection, wang2009observation},
with the discovery of topological crystalline insulators (TCIs) \cite{fu2011topological}, the topological phases were significantly enriched beyond the ten-fold way classification of topological insulators and superconductors \cite{chiu2016classification}. 

However, one has to be cautious when generalizing the ideas from the early examples of topological phases, especially to those that are protected by crystalline symmetries.
For example, due to the fact that crystalline symmetry is often broken at a physical boundary, some TCIs only exhibit robust boundary states at certain crystal orientations \cite{fu2011topological}.
Moreover, with the discovery of novel states such as fragile topological phases \cite{po2018fragile, song2020twisted} and higher order topological insulators (HOTIs) \cite{benalcazar2017quantized, schindler2018higher}, the notion of bulk-boundary correspondence of codimension 1 may not have any direct generalization to TCIs at all.

\begin{figure}
    \centering
    \includegraphics[width = 0.5\textwidth]{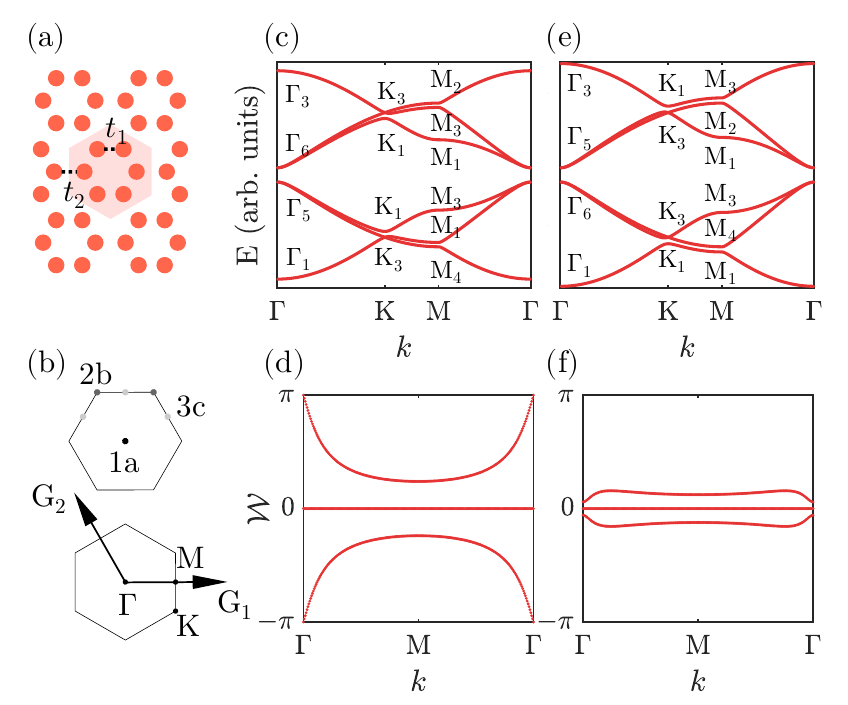}
    \caption{(a) A schematic of the Wu-Hu lattice, the shadowed area indicates the hexagonal unit cell, $t_1$ ($t_2$) correspond to intracell (intercell) couplings.
    When each site moves away from (towards) the unit cell center, $t_1<t_2$ ($t_1>t_2$), and is referred to as an expanded (shrunken) phase.
    Note, when $t_1 = t_2$ this structure corresponds to photonic graphene.
    (b) (top) $1a$, $2b$, $3c$ Wyckoff positions of the unit cell color coded in black, dark gray and light gray, and located at the center, vertices and edges, respectively. (bottom) Brillouin zone of a triangular lattice. 
    (c,d) The band structure of an expanded phase and its corresponding Wilson loop. 
    (e,f) The band structure of a shrunken phase and its corresponding Wilson loop. Irreps are noted in the band diagrams at each high symmetry point.
    Note in (c) and (e), $\Gamma_5$ and $\Gamma_6$ are representations of $d$ and $p$ orbital states, respectively, therefore showing the band inversion.
    In (d) and (f), both phases show trivial Wilson loop without winding from $-\pi$ to $\pi$.}
    \label{fig:fig1}
\end{figure}

The topological photonic crystal proposed by Wu and Hu \cite{wu2015scheme}, which we refer to as the Wu-Hu model, is an elegant structure for realizing a proposed bosonic analog of the fermionic $\mathbb{Z}_2$ TI (Fig. \ref{fig:fig1}). 
Hence it is claimed to host symmetry protected edge states which enable robust light transport free from back-scattering.
The simplicity of the model triggered innumerous experimental and theoretical studies after its discovery \cite{barik2018topological,yang2018visualization, smirnova2019third,shao2020high, dikopoltsev2021topological, liu2020z2,liu2020generation, kumar2022terahertz, zhang2017topological, yang2020spin, liu2017pseudospins, pirie2022topological, cha2018experimental, he2016acoustic, li2021experimental, barik2016two}.
With the development of new theoretical frameworks such as topological quantum chemistry (TQC) \cite{bradlyn2017topological,cano2018building} and the Wilson loop method \cite{alexandradinata2014wilson}, by which one can systematically diagnose nontrivial topology from symmetry representations and Wannier obstruction,
the nature of the bulk topology of Wu-Hu model has been discussed in some recent papers \cite{de2019engineering,palmer2021berry}. 
However, the exact bulk-boundary correspondence has never been identified,
therefore the robustness of the edge states and their relation to the bulk topology remain unclear.
Here we revisit the Wu-Hu model and analyse the nature of its topology with a special emphasis on the edge properties.
Our main result is that, although being topologically distinct, the two phases of Wu-Hu model are trivial in the sense that symmetric localized Wannier functions exist, and both phases do not enforce any protection to the interface states.
We show that the ``topological'' interface states can be drastically gapped even by a uniform symmetric perturbation across the interface, and the associated ``topological'' properties can be well reproduced by trivial defect states.

We briefly review the original formalism of the Wu-Hu model as the foundation of discussion.
The tight-binding model of an expanded or shrunken honeycomb lattice provides a faithful description of Wu-Hu model, in which the unit cell for a graphene lattice is enlarged to include six atomic sites, and the couplings are divided into intra- ($t_1$) and intercell ($t_2$) couplings (Fig. \ref{fig:fig1}a).
When $t_1=t_2$, a four-fold degeneracy appears at the $\Gamma$ point, which gives rise to a ``double Dirac-cone". 
The cell-periodic part of the degenerate Bloch functions have the symmetries of $\ket{p_\pm}$ and $\ket{d_\pm}$ orbitals, and a gap opening and band inversion can be achieved by tuning the relative magnitudes of $t_1$ and $t_2$ (Figs. \ref{fig:fig1}c,e).

At high symmetry momenta, a composite pseudo-fermionic time-reversal symmetry $\mathcal{\tilde{T}}^2=-1$ was constructed and the $\mathbb{Z}_2$ topology was derived through the analogy to the spinful case.
For example, at the $\Gamma$ point in the $(\ket{p_x}, \ket{p_y})$ basis, the pseudo time-reversal operator is given by
\begin{equation}
    \label{eq:pseudo T}
    \mathcal{\tilde{T}}=\mathcal{UK}=[D_{E_1}(C_6)+D_{E_1}(C_6^2)]/\sqrt{3}\cdot \mathcal{K}=-i\sigma_y \mathcal{K}
\end{equation}
in which $E_1$ is the irreducible representation (irrep) for $6mm1'$ (the little co-group at $\Gamma$) furnished by $(\ket{p_x}, \ket{p_y})$ orbitals and $D_{E_1}(C_6)$ is the corresponding matrix representation of $C_6$, $\mathcal{K}$ is the bosonic time-reversal operator functioning as complex conjugate.
Eq.(\ref{eq:pseudo T}) satisfies $\mathcal{\tilde{T}}^2=-1$ and thus protects Kramer's degeneracy at $\Gamma$ point. 
Similar operator were also constructed for $\ket{d}$ orbitals, and for $K$ and $K'$ points in the Brillouin zone (BZ).

The $\mathbb{Z}_2$ index was obtained through the parity of spin-Chern number for each pseudo-spin channel where the $\ket{p_+}(\ket{p_-})$ and $\ket{d_+}(\ket{d_-})$ orbitals are assigned with pseudo-spin up(down) \cite{wu2015scheme, barik2016two}. 
The bulk-boundary correspondence of the 2D spinful TI was directly applied in the original proposal.
The interface states between different phases of Wu-Hu model were claimed to be gapless (with a tiny gap due to the $C_6$ breaking at the interface), immune from back-scattering and possessing spin-momentum locking.

It is however not fully justified why Eq.(\ref{eq:pseudo T}) would constrain the global algebraic classification of Bloch functions and imply physical consequences exactly the same as the time-reversal symmetry in spinful systems.
Here, we examine the topology of the Wu-Hu model based on Wannier obstruction using TQC \cite{bradlyn2017topological,cano2018building} and Wilson loop methods \cite{alexandradinata2014wilson}.
The Wannier obstruction is important because it can be directly related to the topological boundary states \cite{alexandradinata2016topological, fidkowski2011model}.
It has been recently shown that for continuum experimental systems the Wannier obstruction is a necessary condition for the robustness of the interface states \cite{alexandradinata2020crystallographic}, which is of utmost importance.

In TQC theory \cite{bradlyn2017topological,cano2018building}, the symmetry properties of the Bloch functions of Wannier-representable bands is equivalent to a direct sum of elementary band representations (EBRs).
Throughout the BZ, the symmetry properties can be well described by the collection of irreducible representations (irreps) furnished by the Bloch functions for the little groups at high symmetry momenta.
In Figs. \ref{fig:fig1}c,e, we calculate the irreps at high symmetry momenta for both shrunken and expanded phases in the Wu-Hu model 
and the relevant EBRs are listed in Table.\ref{tab:band_rep} \cite{bradlyn2017topological,aroyo2006bilbao1,aroyo2006bilbao2,aroyo2011crystallography}.
For the valence bands (VBs), we obtain $(A_1\uparrow G)_{1a} \oplus (E_1\uparrow G)_{1a}$ for the shrunken case and $(A_1 \uparrow G)_{3c}$ for the expanded case, respectively.
Surprisingly, VBs for both phases transform as a direct sum of EBRs, which suggests the trivial nature of the bulk topology.

We also calculated the phase of the eigenvalues of the Wilson loop operator, which is defined by the following path ordered integral \cite{alexandradinata2014wilson},
 \begin{equation}
     \mathcal{W}_C=\mathcal{P}\exp\left[ i\oint_C \mathbf{A}(\mathbf{k})\cdot d\mathbf{k} \right]
 \end{equation}
where $[\mathbf{A}(\mathbf{k})]_{mn}=i\bra{u_m(\mathbf{k})}\nabla_\mathbf{k}\ket{u_n(\mathbf{k})}$ is the non-Abelian Berry connection for the full VBs.
Fig. \ref{fig:fig1}b shows the geometry of the Wilson loop, where the closed loop $C$ is defined by the reciprocal lattice vector $G_1$ and the spectra is plotted as the loop moves along $G_2$ (Figs. \ref{fig:fig1}d,f).
For both phases of the Wu-Hu model, no winding is observed, which also suggests that the whole VBs can be smoothly deformed into a trivial atomic insulator.


\begin{figure}
    \centering
    \includegraphics[width = 0.47\textwidth]{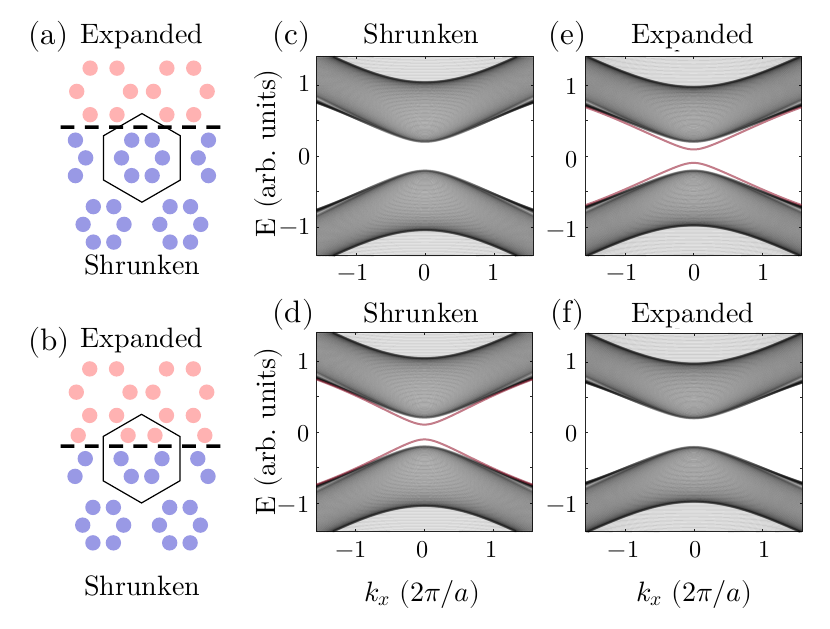}
    \caption{(a),(b) Demonstration of two distinct edge configurations. Red (blue) sites correspond to the expanded (shrunken) phase, and a complete hexagonal unit cell is marked in the figure. In (a), the edge cuts through $3c$ Wyckoff position whereas in (b) through $1a$ Wyckoff position. Note that after cutting, two edge states with open boundary conditions are created in each configuration. (c),(e) Energy dispersion in a strip geometry with edge configuration shown in (a). The gapped edge states only show up in the expanded phase. (d),(f) Energy dispersion with edge configuration shown in (b). The gapped edge states only show up in the shrunken phase.
}
    \label{fig:fig2}
\end{figure}

\begin{table}[b]
\centering
\caption{\label{tab:band_rep} \textbf{The EBRs for space group $P6mm1'$ (the symmetry of the Wu-Hu model).} The EBRs are induced representations of localized orbitals and are labeled by $(\rho\uparrow G)_p$ in which $p$ is the Wyckoff position where the orbitals sit, $\rho$ is the irrep furnished by the orbitals and $G$ is the space group of the system.}
\begin{tabular}{ c c c c }
\hline
\textrm{Band-rep.}&
\textrm{$(A_1 \uparrow G)_{1a}$}&
\multicolumn{1}{c}{\textrm{$(E_1 \uparrow G)_{1a}$}}&
\textrm{$(A_1 \uparrow G)_{3c}$}\\
\hline
$\Gamma$ & $\Gamma_{1}$ & {$\Gamma_6$} & $\Gamma_1 \oplus \Gamma_5$\\
$K$ & $K_1$ & {$K_3$} & $K_1\oplus K_3$\\
$M$ & $M_1$ & $M_{3}\oplus M_{4}$ & $M_{1}\oplus M_{3}\oplus M_{4}$\\
\hline
\end{tabular}
\end{table}

Next, we briefly discuss the topological invariants for the VBs of the Wu-Hu model.
In Supplementary materials \cite{supp} we prove that the spin-Chern number and the $\mathbb{Z}_2$ index defined for Wu-Hu model are equivalent to the Euler class and the second Stiefel-Whitney class protected by $C_2\mathcal{T}$ symmetry \cite{ahn2019stiefel}.
In 2D systems with $C_2\mathcal{T}$ symmetry, the Euler class is a $\mathbb{Z}$ type fragile topological invariant defined in a two-band subspace \cite{ahn2019stiefel, cano2021band}.
A non-zero Euler class forbids the construction of symmetric localized Wannier functions, however, this obstruction may be lifted by adding trivial bands.
In this many-band limit, the parity of the Euler class becomes the well defined $\mathbb{Z}_2$ type second Stiefel-Whitney class, $\omega_2$.
The expanded phase belongs to this category and is characterized by a non-trivial $\omega_2=1$ which indicates an obstructed atomic limit (OAL), meaning that the Wannier functions cannot be localized at the center of the unit cell. 
The associated physical consequence is a quantized quadrupole moment and fractional corner charges, in other words, $\omega_2=1$ characterizes a HOTI \cite{ahn2019failure, ahn2019stiefel}.

In fact, the VBs for both phases of Wu-Hu model can be deformed into decoupled atomic clusters by selectively turning off intra- or inter-cell couplings. 
With all the above observations we conclude that both phases of the Wu-Hu model are topologically trivial in terms of Wannier obstruction, therefore neither of the two phases are responsible for the gapless interface states.
This can be easily demonstrated in the tight binding limit.
Starting with the gapless interface and adiabatically turning off the couplings connecting two phases to form the open boundary conditions (OBCs), the interface states would be in general gapped and pushed towards the bulk bands. 
If the interface states result from the nontrivial bulk topology, we can keep track of them and the gapped interface states should be localized exactly at the non-trivial half of the system.
However, depending on the edge configuration, the interface states of Wu-Hu model can be localized at different phases.
In Fig. \ref{fig:fig2}, we show that
for the shrunken phase where the Wannier center sits at $1a$ Wyckoff position, the gapped edge states appear when the boundary cuts through $1a$ position; for the expanded case where the Wannier center sits at $3c$ Wyckoff position, the gapped edge states appear when the edge cuts through $3c$ Wyckoff position.
This observation strongly suggests that the interface states are originated from the local defects in contrast to the well-known topological boundary states arising from the bulk Wannier obstruction \cite{alexandradinata2016topological, fidkowski2011model}. 



The Kramer's degeneracy in 1D BZ cannot be protected by composite pseudo time-reversal symmetry in the Wu-Hu model hence invalidating the direct generalization from the 2D spinful TI protected by $\mathcal{T}^2=-1$.
An alternate interpretation explains the interface states as the Jackiw-Rebbi soliton eigen-solutions that arise from a local band inversion \cite{barik2016two}.
However, since the Jackiw-Rebbi solutions give one set of interface states for each pseudo-spin, spin-mixing can potentially gap out the interface states, and the symmetry that protects the bulk topology in the Wu-Hu model, namely $C_6$ and $\mathcal{T}^2=1$, does not imply spin-conservation.
Consider the Wu-Hu model in its quasi-orbital basis, where $\ket{p_\pm}$ and $\ket{d_\pm}$ orbitals sit at 1a Wyckoff position of a triangular lattice. 
The spin flipping terms are locally forbidden by $C_6$ symmetry since different pseudo-spin orbitals belong to irreps with different $C_6$ eigenvalues.
However, the following non-local spin-flip channel is always allowed by $C_6$ symmetry,
\begin{equation}
    \Delta=t a^\dag_{i,\pm}a_{j,\mp} + h.c., \ i\neq j
\end{equation}
where $i,j$ are labels of unit cells, $\pm$ are labels for pseudo-spin, and $h.c.$ stands for hermitian conjugate.

\begin{figure}
    \centering
    \includegraphics[width = 0.4\textwidth]{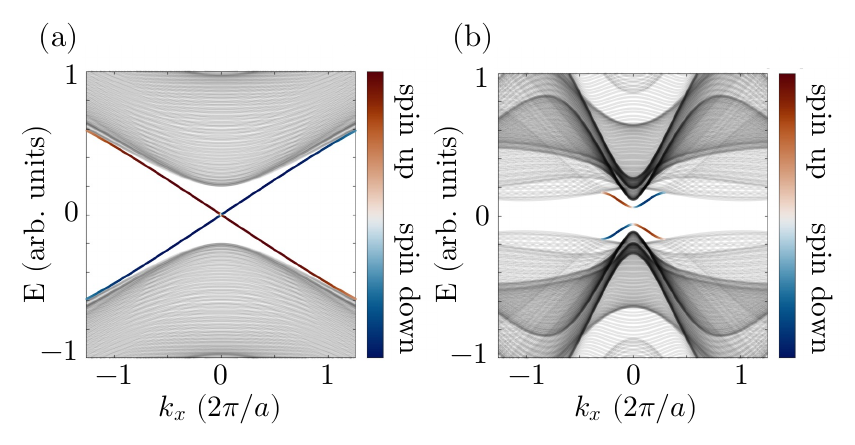}
    \caption{The dispersion of interface states where the pseudo-spin component is color coded. (a) The interface states of an unperturbed Wu-Hu interface. The dispersion is nearly linear and the gap is not visible in the figure. (b) Perturbed Wu-Hu interface. An apparent gap is opened with magnitude comparable to the bulk band gap. The pseudo-spins are mixed showing lighter color. $a$ is the lattice constant.}
    \label{fig:fig3}
\end{figure}

To further demonstrate the lack of robustness in the Wu-Hu interface, we explicitly show that they can be gapped considerably even by a $C_6$ and $\mathcal{T}^2=1$ symmetric perturbation uniform across the interface (Fig. \ref{fig:fig3}).
The perturbation is added to ensure that when $t_1=t_2$, a double Dirac-cone appears at $\Gamma$ point.
The band inversion is then achieved by tuning the relative magnitude of $t_1$ and $t_2$ (see Supplementary materials \cite{supp}).
Therefore the original Wu-Hu Hamiltonian is explicitly included and the perturbation is uniform across the interface preserving $C_6$ symmetry. 
Importantly, this perturbation can be viewed as being adiabatically added to the unperturbed Hamiltonian and no gap closing ever happened between the VBs and CBs, thus the topology is preserved.
For a system with an interface, we write the perturbed Hamiltonian as:
\begin{equation}
    H'=H_0 + \Delta H
\end{equation}
in which $H_0$ describes the unperturbed interface of the Wu-Hu model and $\Delta H$ is the perturbation.
The spectra of interface states for $H'$ and $H_0$ are shown in Fig. \ref{fig:fig3}.
For $H_0$, there exists a gap at zero energy that is hardly visible (as observed in the Wu-Hu model \cite{wu2015scheme}), whereas for $H'$, the gap is clear and considerably large compared to the bulk band gap.
Pseudo-spin character of the interface states also shows clear mixing for $H'$ compared to $H_0$, which is consistent with the argument that $C_6$ symmetry does not imply spin-conservation.
All these observations strongly suggests that, aside from $C_6$ symmetry breaking, other mechanisms can open a gap for the interface states, therefore showing the absence of topological protection in the system clearly.

In addition, the Wu-Hu interface and the edge of 2D TIs protected by $\mathcal{T}^2=-1$ are compared to discuss the relation between their properties and topology.
The three properties concerned here are spectral robustness, immunity from back-scattering and spin-momentum locking.
For the $\mathcal{T}^2=-1$ 2D TI, the bulk-boundary correspondence can be understood by the topological equivalence between the edge spectrum and the Wilson loop spectrum, which has a stable winding protected by Wannier obstruction in the non-trivial phase, so that the spectral robustness of the edge is guaranteed \cite{alexandradinata2016topological, fidkowski2011model}.
The immunity of back-scattering is then followed as a combined effect of $\mathcal{T}^2=-1$ and the presence of odd number of edge states \cite{kane2005quantum}.
Lastly, instead of a unique topological phenomenon, the spin-momentum locking is a prevalent feature in edge modes with strong spin-orbit coupling.
To conclude, only spectral robustness is directly related to topology.
Moreover, in bosonic systems with $\mathcal{T}^2=1$ symmetry, back-scattering can still be present even in edge states with topologically protected spectral robustness.
For Wu-Hu interface states, this argument agrees well with recent experimental results \cite{orazbayev2019quantitative}.

From a practical perspective, these gapless, back-scattering free and spin-momentum locking interface states are what make the Wu-Hu model promising for photonic applications.
Among which only spectral robustness is possible to be directly linked with topology. 
The unnecessity of topology at the Wu-Hu interface enables more flexible designs of combining different bulk structures without any symmetry consideration, which may lead to novel applications such as photonic on-chip logic and reconfigurable light routing.
Here we numerically demonstrate in a 2D photonic crystal helical edge states that solely stem from the trivial phase of Wu-Hu model with OBCs that reproduces all the features of the claimed ``topological'' Wu-Hu interface. Structures and corresponding bulk band diagrams can be found in Supplementary materials \cite{supp}.

\begin{figure}
    \centering
    \includegraphics[width = 0.5\textwidth]{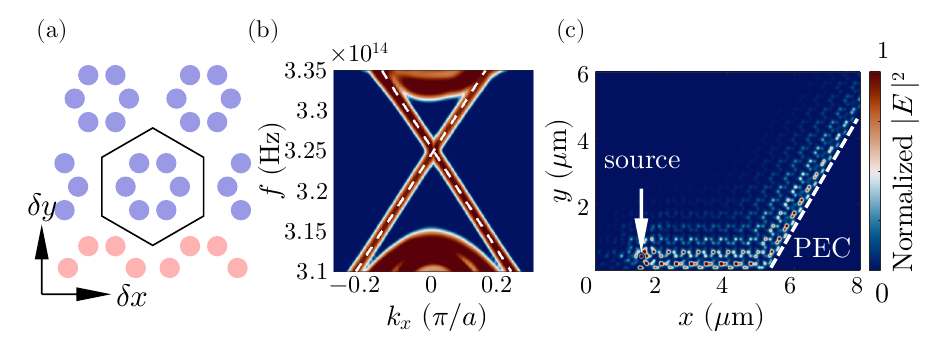}
    \caption{(a) The schematic of the strip geometry applied for the edge states of a shrunken phase, atoms from bulk complete (edge incomplete) unit cell are colored in blue (red). The edge is created by cutting through the 1a position, then a slight tuning is applied to the incomplete unit cells at the edge. The direction of $\delta x,\ \delta y$ is also noted.  (b) Numerically calculated interface dispersion, showing two in-gap linear modes. $a$ is the lattice constant (c) Large scale simulation of the propagation of a trivial edge state from a circularly polarized source. The open boundary turning is marked in a white dashed line.
}
    \label{fig:fig4}
\end{figure}

Edge states are created from the shrunken phase by cutting through the Wannier center of the VBs, namely $1a$ Wyckoff position, forming incomplete unit cells. (Fig. \ref{fig:fig4}a)
Being of the defect nature, the resulted edge states are highly tunable that they can be tuned to be gapless by simply displacing the sites in the incomplete unit cells.
We first calculated the dispersion spectrum of a strip geometry of this trivial edge (Figs. \ref{fig:fig4}a, b), with Bloch boundary condition applied in the x-direction and OBCs in the y-directions (see Supplementary materials \cite{supp} for detailed simulation setup).
Two edge states emerge in the dispersion inside the bulk gap (Fig. \ref{fig:fig4}b), showing a dirac-cone shaped crossing.
Then we performed a large scale simulation of the edge states with a sharp bend excited with a circularly polarized source (Fig. \ref{fig:fig4}c).
The unidirectional propagation is clearly observed along the sample edge, showing that topology is not required for a helical photonic edge.

In conclusion, we re-examined the Wu-Hu model and identified the algebraic nature of the topological invariants and the associated physical consequences.
We showed the lack of robustness of its interface states against symmetry preserving perturbations and explicitly constructed a trivial defect edge that reproduces all the ``topological'' properties. 
However, the following question remains interesting and unanswered: for TCIs, whether, and to what extent, Wannier obstructions would provide protection to the the interface in the domain wall configuration similar to the Wu-Hu model where the bulk symmetry is partially restored by the addition of a trivial phase.
In fact, the existence of such topological protection is an implicit assumption for the topological interpretation of Wu-Hu interface.
If this protection does not exist even when one of the phases is stably obstructed, the topological interpretation of Wu-Hu interface would fail at the first step.
Based on our arguments, one cannot distinguish whether the trivial nature of the VBs or the absence of topological protection itself is the fundamental reason that is responsible for the gap opening.
The rigorous discussions of similar questions has only appeared recently \cite{alexandradinata2020crystallographic}, and we hope our results as a case study can provide some insights to future studies.

\begin{acknowledgments}
This work was supported by the Office of Naval Research via grant No.N00014-22-1-2378. S.X. and Y.W. contributed equally to this work.
\end{acknowledgments}


%

\end{document}



\title[Supplementary materials]{Supplementary materials for:\\Absence of topological protection of the interface states in $\mathbb{Z}_2$ photonic crystals}

\author{Shupeng Xu}
\author{Yuhui Wang}%
\author{Ritesh Agarwal}
\email{riteshag@seas.upenn.edu}
\affiliation{
Department of Materials Science and Engineering, University of Pennsylvania,\\Philadelphia, 19104, PA, US}
\date{\today}

\maketitle

\section{\label{sec:index} Topological Indices of the Wu-Hu model}
Here we review how the topological indices are obtained in the original proposal and prove their equivalence to the Euler class and the secon Stiefel-Whitney class.
In the tight binding model, near the critical point of the phase transition ($t_1=t_2$), the $k\cdot p$ expansion at $\Gamma$ point around fermi energy gives a well defined change in spin-Chern number.
Equivalently the spin-Chern number can be directly calculated for the projected Hamiltonian in $(\ket{p+}, \ket{d+}, \ket{p-}, \ket{d-})^T$ basis.
\begin{Sequation}
    \label{eq:projected}
    H_p = \mathcal{P}H_0\mathcal{P} = \begin{pmatrix}
    H_+ & G\\
    G^\dag & H_-
    \end{pmatrix} 
\end{Sequation}
Where $H_0$ is the full Hamiltonian, $\mathcal{P}$ is the projector and $G$ is the off-diagonal matrix. 
We further define a spin-conserved Hamiltonian with zero off-diagonal blocks
\begin{Sequation}
    \label{eq:kp}
    H_{spin} = \begin{pmatrix}
    H_+ & 0\\
    0 & H_-
    \end{pmatrix}
\end{Sequation}
and a pseudo-spin gauge that satisfies
\begin{Sequation}
    C_2\mathcal{T}\ket{u_{+,n\mathbf{k}}}=\ket{u_{-,n\mathbf{k}}}
\end{Sequation}
where $i=\pm$ is the label of pseudo-spin,  $\ket{u_{\pm, n\mathbf{k}}}$ are eigenstates of $H_{\pm}$ and $n=1,2$ is the index of eigenstates within each block.
The non-Abelian Berry curvature $\mathbf{F}_{mn}=\nabla_{\mathbf{k}}\times\mathbf{A}_{mn}$ ($\mathbf{A}_{mn}=\bra{u_{m\mathbf{k}}} \nabla_{\mathbf{k}} \ket{u_{n\mathbf{k}}} $ is the non-Abelian Berry connection) for the lower two bands in the pseudo-spin gauge is diagonal

\begin{Sequation}
    F_z= \begin{pmatrix}
    i\Omega_+ & 0 \\
    0 & i\Omega_-
    \end{pmatrix}
\end{Sequation}

in which $\Omega_\pm$ are real numbers and $\Omega_+=-\Omega_-$ due to $C_2\mathcal{T}$ symmetry. 
The spin-Chern number $C_\pm$ is the integral of $\Omega_\pm$ over BZ and satisfies $C_\pm=0$ and $C_\pm=\pm1$ for the shrunken and expanded case respectively.

In a two-band subspace of a 2D $C_2\mathcal{T}$ symmetric system the Euler class can be expressed as a flux integral in the real gauge which is defined as
\begin{Sequation}
    C_2\mathcal{T}\ket{\tilde{u}_{n\mathbf{k}}} = \ket{\tilde{u}_{n\mathbf{k}}}
\end{Sequation}

In the real gauge the Berry connection $\tilde{\mathbf{A}}_{mn}$ and Berry curvature $\mathbf{\tilde{F}}_{mn}$ are 2D real anti-symmetric matrices, so they only have one degree of freedom. 
The Euler class can be written as the integral of the off-diagonal element
\begin{Sequation}
    e_2 = \frac{1}{2\pi}\oint_{BZ} d\mathbf{S}\cdot \mathbf{\tilde{F}}_{12} \in \mathbb{Z}
\end{Sequation}
We note that the real gauge is connected to the pseudo-spin gauge by the following gauge transformation
\begin{Sequation}
    \ket{\tilde{u}_{i,n\mathbf{k}}} = \sum_{j}U_{ij}\ket{u_{j,n\mathbf{k}}}
\end{Sequation}
where
\begin{Sequation}
    U=\frac{1}{\sqrt{2}}\begin{pmatrix}
    1 & 1 \\
    i & -i
    \end{pmatrix}.
\end{Sequation}
And the Berry-curvature transforms as
\begin{Sequation}
    \tilde{F}_z=U^*F_zU^T=\begin{pmatrix}
    0 & \Omega_- \\
    \Omega_+ & 0
    \end{pmatrix}.
\end{Sequation}
Thus we demonstrated the equivalence between the spin-Chern number and the Euler class for $H_{spin}$ with $e_2=C_{\pm}=0$ and $\pm e_2=C_{\mp}=\mp1$ for the shrunken and expanded case of Wu-Hu model respectively.

The Euler class remains invariant for the projected Hamiltonian $H_P$ since it can be obtained from Eqn.\ref{eq:kp} through a symmetric topology conserving deformation.
We note that the fragile nature of non-trivial Euler class agrees with the EBR analysis for projected bands in the expanded case.
The bands can be written as the difference of two EBRs: $\mathcal{B}=(A_1\uparrow G)_{3c}\ominus(A_1\uparrow G)_{1a}$.

The full valence band of Wu-Hu model can be viewed as the direct sum of the projected bands and a trivial band $(A_1\uparrow G)_{1a}$, and the parity of Euler class survives as the second Stiefel-Whitney class $\omega_2$ where $\omega_2=0$ and $\omega_2=1$ correspond to the shrunken and expanded phases, respectively.

Lastly, we noted that with the presence of $C_2$ symmetry, $\omega_2$ can be expressed using the product of $C_2$ eigenvalues of Bloch states:
\begin{Sequation}
\label{eq:inversion eigvl}
    (-1)^{\omega_2}=\prod_{i=1}^{4}(-1)^{\lfloor N_{occ}^-(\Gamma_i)/2 \rfloor}
\end{Sequation}

where $\Gamma_{i=1,2,3,4}$ are four $C_2$ invariant momenta in 2D BZ, $N_{occ}^-(\Gamma_i)$ is the number of occupied bands with negative $C_2$ eigenvalues at $\Gamma_i$ and $\lfloor\ \rfloor$ is the floor function or the greatest integer function.
This simple expression exactly coincidences with the topological index in Ref.\cite{liu2019helical}.

\section{Details of Perturbed Wu-Hu Model}
The perturbation is added as long range couplings shown in Fig. \ref{fig:figApp} and all its $C_6$ partners.
The hopping parameters are chosen as $l_1 = 0.28$, $l_2 = 0.4$, $l_3 = 0.68$, $l_4 = 0.4$ and $l_5 = 0.28$.
The hopping strength satisfies $l_1=l_5$ and $l_2=l_4$ which is required by $C_6$ symmetry.
Further considering the requirement described in the main text that a Dirac cone appears when the intra- and inter- cell couplings equal to each other, there are only two independent degrees of freedom for the perturbation.
The magnitudes for the original couplings of two phases of Wu-Hu model are $t_1 = -1.1$, $t_2 = -0.9$ (shrunken) and $t_1 = -0.9$, $t_2 = -1.1$ (expanded) respectively.

\renewcommand{\thefigure}{S1}
\begin{figure}
    \centering
    \includegraphics[width = 0.25\textwidth]{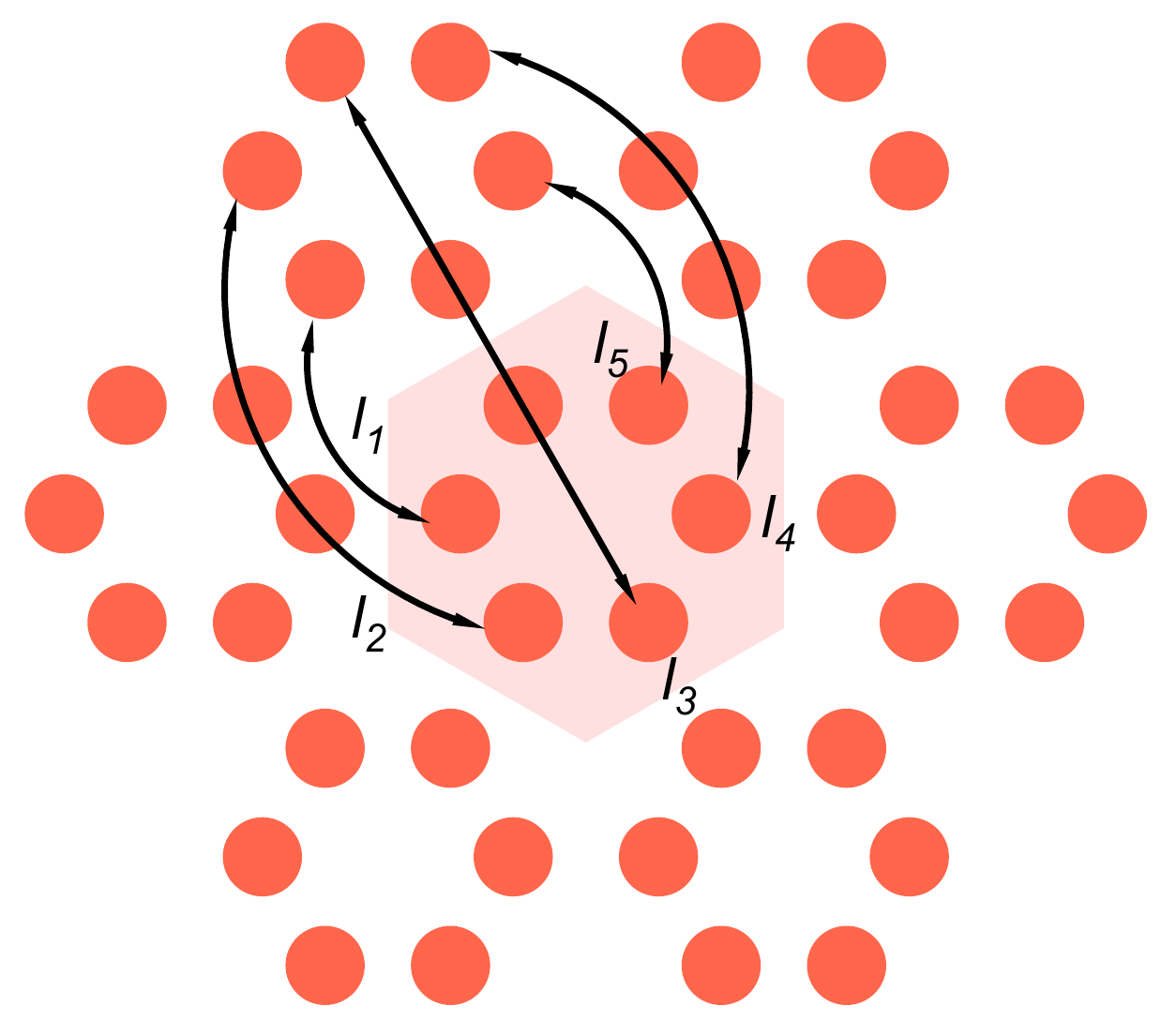}
    \caption{A schematic of one copy of the long-rang coupling for the uniform symmetric perturbation applied to the Wu-Hu model, Fig. 3 in the main text. To preserve $C_6$ symmetry, this set of couplings is accompanied with all its $C_6$ partners.}
    \label{fig:figApp}
\end{figure}

\section{\label{app:sim} Numerical simulations of trivial helical edge states}
\renewcommand{\thefigure}{S2}
\begin{figure}
    \centering
    \includegraphics[width = 0.48\textwidth]{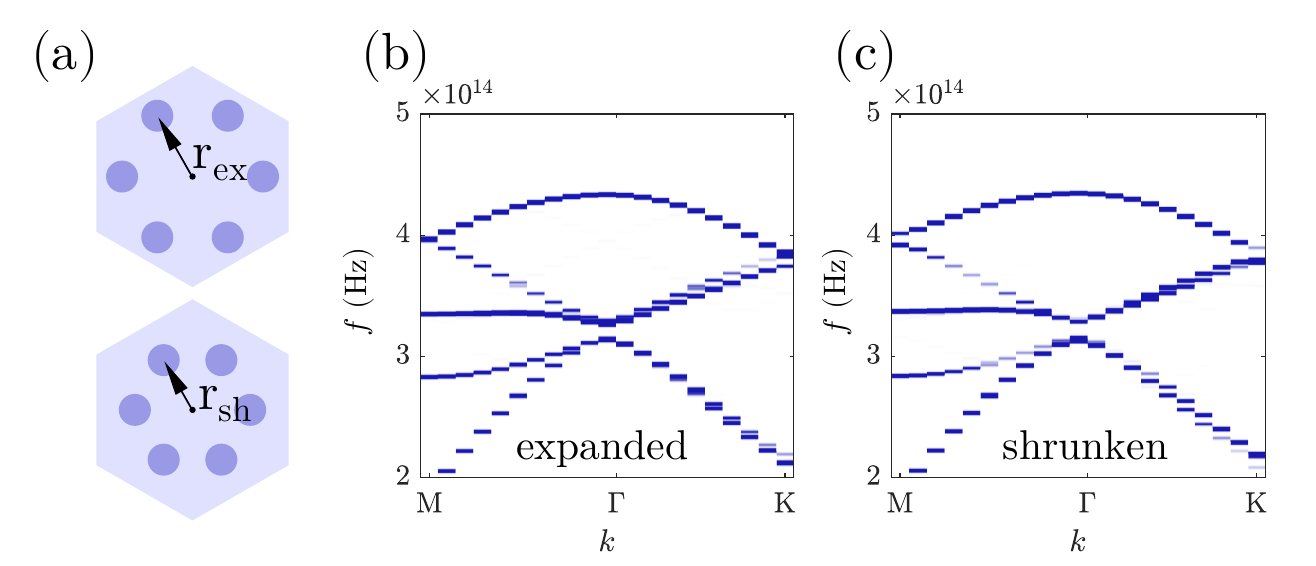}
    \caption{(a) Schematics of the expanded (top) and shrunken (down) unit cells for the original Wu-Hu model used in numerical simulations, respectively. $r_\text{ex}$ and $r_\text{sh}$ are the distances from each site to the center of the unit-cell, marked in the figure. (b) and (c) FDTD simulation results of the bulk band structure of the expanded and shrunken phases of the original Wu-Hu model used in the main text.}
    \label{fig:figApp2}
\end{figure}
We used a commercial finite-difference time-domain (FDTD) software package (\textsf{Lumerical FDTD}) for the electro-magnetic field numerical simulation to verify our conclusions that nontrivial topology is not required for the helical edge states.
The 2D simulations for the reduction of computation cost are set in a unified design of the dielectric honeycomb lattice: the lattice constant is 450 nm, and for each site we set its radius to be 50 nm and corresponding $\varepsilon = 11.7$. All simulations were conducted in a vacuum background. For the graphene structure, the distance between each site and the center of the hexagonal unit cell, defined by $r_0$, was $150$ nm. For the shrunken phase $r_{\text{sh}} = 142.9$ nm and for the expanded phase $r_{\text{ex}} = 155.2$ nm. (Fig. \ref{fig:figApp2}a) 

We presented the results for the band dispersion calculations of shrunken and expanded phases, shown in Figs. \ref{fig:figApp2}b and c, as a general reference for the simulations in the main text. 
A complete gap can be noticed near 320 THz.
For the simulation of the edge states, Bloch boundary conditions were defined in the x direction, with a perfect electrical conductor (PEC) applied at the bottom and a perfect matched layer (PML) in the top of the y direction, respectively. 
The positions for the bottom two sites in the incomplete unit cells were tuned by $\delta x = \pm 30$ nm to move away from the corresponding center of the unitcell, and $ \delta y = 10 $ nm upwards (Fig. 4a). The line of the incomplete unit cells (red sites in Fig. 4a) was then brought close to bulk part by $\delta y_{\mathrm{line}} = 40$ nm.

For the large scale propagation, a set of circularly polarized magnetic dipoles were placed at the center of a complete unit cell near the edge of the system. The excitation frequency was set at 322.8 THz.
The bottom and right boundaries (which defined the 120-turning) were set to be PEC and metal, functioning as OBCs. The intensity is normalized to the peak intensity outside the source region.

%
